# L shell x-ray production in high-Z elements using 4-6 MeV/u fluorine ions

Sunil Kumar[1,2], Udai Singh[2], M. Oswal[3], G. Singh[1], N. Singh[1], D. Mehta[1], G. Lapicki[4] and T. Nandi[5]*

[1]*Department of Physics, Panjab University, Chandigarh160014, India.*

[2]*Department of Applied Sciences, Chitkara University, Himachal Pradesh 174103, India.*

[3] *Department of Physics, Dev Samaj College Sec 45-B Chandigarh 160047, India*

[4]*Department of Physics, East Carolina University, Greenville, North Carolina 27858, USA.*

[5]*Inter-University Accelerator Centre, New Delhi110067, India.*

## Abstract

L shell line and total x-ray production cross sections in $_{78}$Pt, $_{79}$Au, $_{82}$Pb, $_{83}$Bi, $_{90}$Th, and $_{92}$U targets ionized by 4-6 MeV/u fluorine ions were measured. These cross sections are compared with available theories for L shell ionization using single- and multiple-hole fluorescence and the Coster-Kronig yields. The ECPSSR and the ECUSAR theories exhibit good agreement with the measured data, whereas, the FBA theory overestimates them by a factor of two. Although for the F ion charge states q = 6-8 the multiple-hole atomic parameters do not significantly differ from the single-hole values, after an account for the multiple-holes, our data are better in agreement with the ECUSAR than the ECPSSR theory.

**Keywords**: L-shell x-rays, heavy ions, ionization, multiple ionization.

PACS number(s): 34.50.Fa, 34.80.Dp, 31.15.xp, 32.30.Rj

# 1. INTRODUCTION

The measurement of emitted x-rays from targets has resulted in major advances in radiation[1], plasma [2], atomic and nuclear physics [3], and in particle induced x-ray emission (PIXE) technique [4,5]. While PIXE originated and continues using light ions such as protons or alphas [6–16],there is an increasing interest to use heavy ions for PIXE analysis due to higher cross sections and thereby better sensitivity [17]. While discrepancies between theories and experiment were attributed to multiple ionization even with protons[18], multiple-ionization effect has been known for decades in L-shell ionization by heavier ions [19–34]. However this effect is still rarely addressed for the x-ray emission elemental analysis in the aftermath of ionization by such ions.

The sum of electron capture (EC) from a projectile with the atomic number $Z_P$ and direct ionization (DI) of a target with the atomic number $Z_T$ results in ionization of the target atom's inner shells. In asymmetric collisions, i.e., $Z_P/Z_T \ll 1$, the DI is dominant, whereas, for symmetric collisions, i.e., with $Z_P/Z_T$ approaching 1, the EC process becomes increasingly important. As presented in Section 2, the L shell x-ray production cross sections have been measured in high $Z_T$-targets ionized by the 76–114 MeV $^{19}$F ions. With $Z_P = 9$, $0.010 \leq Z_P/Z_T \leq 0.012$ and the ratio of the projectile velocity $v_p = 6.351 [E_P(MeV)/A_P(u)]^{1/2}$ (a.u) to the orbital velocity of the L-shell electrons $v_T=(Z_T-4.15)/2$ less than 1 i.e., $0.029 \leq v_P/v_T \leq 0.042$, the present data are in the asymmetric and slow collision regime.

While expanding on the existing data base with ionization by heavy ions as desired for PIXE analysis, the collision regime of the present data allows for a meaningful comparison with existing ionization theories as discussed in Section 3. Section 4 addresses effects of the single- and multiple-hole atomic parameters required for conversion of ionization to x-ray production cross sections, and Section 5 summarizes our findings.

# 2. EXPERIMENTAL DETAILS AND DATA ANALYSIS

The L shell x-ray production cross sections in the elements with $78 \leq Z_T \leq 92$ elements using the $^{19}$F ions (charge states q = $6^+$, $7^+$, $8^+$) in the 76 – 114 MeV energy range had been measured. Heavy ions of $F^{6+}$ (76 and 84 MeV), $F^{7+}$ (90 MeV) and $F^{8+}$ (98, 106 and 114 MeV) were obtained from the 15 UD Pelletron accelerator at Inter-University Accelerator Centre,

New Delhi. Two silicon surface barrier detectors at ± 7.5° to the beam direction were used to monitor the projectile ions. The chamber was evacuated to about $10^{-6}$ Torr and equipped with a 5 mm diameter collimator and 6 $\mu$m Mylar window in front of the Si(Li) detector. In the energy range of the measured L x-ray spectra, the energy resolution of the detector was ~200 eV for the Mn K$\alpha$ x rays. A Si(Li) solid state detector (thickness = 5 mm, diameter = 10 mm, 25 $\mu$m Be window from ORTEC, Oak Ridge, Tennessee, USA) was placed in the horizontal ion beam plane configuration outside the vacuum chamber at an angle of 125° to the beam direction and a distance of 170 mm from the target. The targets were mounted on a steel ladder at a 90° angle to the beam direction. The ladder could accommodate up to 24 targets (8 rows and 3 columns) each of 11.7 mm diameter and the desired target was brought along the beam direction by the horizontal and the vertical movement of the target ladder using the stepper motor arrangement. The spot size of the ion beam at the target was ~ 2 mm diameter. The spectra were taken at different positions of each target by tiny steering the beam. The thickness and the uniformity of these targets were measured by the energy loss method using alpha particles from a radioactive decay of $^{241}$Am. Targets of $_{78}$Pt, $_{79}$Au, $_{82}$Pb, and $_{83}$Bi (thickness ~ 120 $\mu$g/cm$^2$) were prepared on the 20$\mu$g/cm$^2$ carbon backing using the vacuum deposition technique [35]. Thinner and spectroscopically pure (99.999 % pure) targets of ThF$_4$ (48.7 $\mu$g/cm$^2$) and UF$_4$ (48.6 $\mu$g/cm$^2$) on Mylar backing (thickness = 3 $\mu$m) procured from Micromatter, Deer Harbor, Washington, USA were also used in the present work. The target uniformity was verified to be better than 5%. The beam current was kept below 1nA to avoid the pile up effects and the damage to the target. The spectra were collected for 30 minutes to 1 hour so as to get good statistical accuracy.

Figure 1 shows typical L x-ray spectra from the targets of $_{78}$Pt, $_{79}$Au, $_{82}$Pb, $_{83}$Bi, $_{90}$Th, and $_{92}$U elements ionized by the 98 MeV $^{19}$F ions. These spectra result from ionization of L$_i$ (i = 1-3) subshells, with which x-ray line peaks correlate viz., Ll, L$\alpha_{1,2}$, and L$\beta_{2,15,6,7}$ from the L$_3$ subshell, the L$\eta$, L$\beta_1$, and L$\gamma_{1,5}$ from the L$_2$ subshell, and the L$\beta_{3,4}$ and L$\gamma_{2,3,4}$ from the L$_1$ subshell. Figure 2 displays L x-ray spectra of $_{92}$U target bombarded by the $^{19}$F ion beam at different energies. The differential L x-ray production cross sections for the major peaks were evaluated with

$$\frac{d\sigma_i^x}{d\Omega} = \frac{N_x A}{N_A N_P t \varepsilon \beta} \qquad (1)$$

where $N_x$ is the net x-ray counts per second under the L x-ray peak, A is atomic mass (in grams), $N_A$ is the Avogadro's number, and $N_p$ is the number of incident ions collected in the Faraday cup. The ion beam changes its charge state during its passage through the target. The mean distribution in charge state of ion beam after passing through the target and its backing is calculated using the computer code ETACHA [36]. This code accounts for electron loss, capture, and excitation from and to all the subshells based on an independent electron model. The measured charge in the Faraday cup using a current integrator has been corrected for the change in the charge state and used for $N_p$ with the incident charge state in Equation (1). Also in this equation, t is the target thickness in $\mu g/cm^2$, $\varepsilon$ is the absolute detection efficiency (included all absorbing components of the set-up), and $\beta \equiv [1 - \exp(-\mu t)]/\mu t$ is the correction factor for the absorption of the emitted L x-rays in the present target, where $\mu$ in $cm^2/\mu g$ is the attenuation coefficient [37]. The $\beta$ is $\geq 0.99$ for the target thickness used in the present measurements. The energy loss calculation using the SRIM code [38] for the incident beam within the target suggests negligibly small energy loss for the target thickness and the beam energies used in the present work. For example, 76 and 114 MeV fluorine ion lose 267 and 223 keV in Pt, and 104 and 87 keV in U target, respectively. The peak areas, $N_x$, are evaluated using the computer program CANDLE [39]. This software is an improved version of the Levenburg-Marquardt [40] non-linear minimization algorithms for the peak fitting. The FWHM for the intrinsic Lorentzian broadening associated with the L x-ray lines is < 12 eV [41]. The energy calibration of the detector is performed before and after the in-beam measurements. Relative efficiency of the x-ray detector in the energy region of interest is deduced by measuring the fluorescence K x-ray yields from various elemental targets excited by the 59.54 keV photons from a point 100 mCi $^{241}$Am source, which was mounted in the chamber instead of the ion beam. The Cu-Al attenuator of suitable thickness is used with the source to remove the low energy $_{93}$Np L x-rays and 26 keV $\gamma$- ray emitted from the source. Thick targets of $_{26}$Fe, $_{28}$Ni, $_{29}$Cu, $_{30}$Zn, $_{33}$As, $_{34}$Se, $_{39}$Y, $_{40}$Zr, $_{41}$Nb, $_{42}$Mo, $_{46}$Pd, $_{47}$Ag, $_{48}$Cd, $_{49}$In, and $_{50}$Sn elements were excited by the 59.54 keV photons. The efficiency of the detector is calculated from

$$I_0 \Omega \varepsilon_{KX} = \frac{4\pi N_{KX}}{\sigma_{KX} t \beta_{KX}} \qquad (2)$$

where $I_o$ is the intensity of the incident photons to be collected into the solid angle $\Omega$ and $\varepsilon_{KX}$ is the absorption of the x-rays in air and Mylar window. $N_{KX}$ is the measured count rate under the K x-ray peak, while $\sigma_{KX}$ is calculated as product of the K shell photoionization cross section [42] the fluorescence yield [43,44], and the fractional emission rates [45] for the Kα and Kβ x-rays. As in Eq.(1), t is the thickness of target element and $\beta_{KX}$ is the absorption correction factor that now depends both on the incident $\theta_i$ and emitted angles $\theta_e$ with respect to the normal to the target. This self-absorption correction factor, accounting for the attenuation of the incident and the emitted K x-rays of the target element, is given by

$$\beta_{KX} = \frac{1 - \exp[-(\mu_i/\cos\theta_i + \mu_e/\cos\theta_e)t]}{(\mu_i/\cos\theta_i + \mu_e/\cos\theta_e)t} \quad (3)$$

where, $\mu_i$ and $\mu_e$ are the mass-attenuation coefficients for the incident and the emitted x-rays in the target calculated using XCOM [37]. A semi-empirically fitted relative efficiency curve is generated by taking into account the absorption of the various K x-rays of the target elements. Figure 3 shows the absolute efficiency $\varepsilon$ of the Si(Li) detector obtained using the calibrated radioactive sources of $^{137}$Cs and $^{155}$Eu. The relative efficiency curve obtained by measuring the K x-ray yields is normalized with respect to the absolute source strength to obtain an absolute efficiency curve.

Although the Ll line that is weakest in L x-ray spectra is not perfectly isotropic [46–48] differential x-ray production cross sections have been measured at an emission angle $\Psi = 125°$ where the second-order Legendre polynomial term, $P_2(\cos\Psi) \approx 0$. Thus integrated x-ray production cross sections were deduced by multiplying the differential cross sections of Eq.(1) by a factor of $4\pi$.

The percentage error in the measured x-ray production cross sections is about 10-15%. This error is attributed to the uncertainties in different parameters used in the analysis, namely, the photopeak area evaluation (~ 5% for the Ll x-ray peak and 3% for the other peaks), ion beam current (~ 7%), target thickness (~ 3%). The error in the absolute efficiency values, $\varepsilon$, is 5-8% in the energy region of interest. The measured cross sections taken for an element from different locations on the same target are found to agree within the experimental error and their weighted average is given in Table 1.

## 3. IONIZATION THEORIES

Direct ionization (DI) of inner shells can be calculated with the plane wave Born approximation (PWBA) [49–51], binary encounter approximation (BEA) [7,52] and semi-classical approximation (SCA) [53] while the Oppenheimer-Brinkman-Kramers formulation of Nikolaev (OBKN) [54] may be used to evaluate electron capture (EC). The sum of the PWBA and OBKN constitute the first Born approximation (FBA) for inner-shell ionization [49–51,54] calculations. An approach that goes beyond the FBA is the ECPSSR theory that accounts the energy-loss (E) and Coulomb-deflection (C) of the projectile and perturbed-stationary state (PSS) and relativistic (R) nature of the target's inner shells [55]. PSS formulas of the ECPSSR theory were further modified for united and separated atom (USA) treatments of the electron wave function to generate the ECUSAR theory [56].

The theoretical x-ray production cross sections $\sigma_{Lp}^{x}(p=l, \alpha, \beta, \gamma)$ for the most commonly resolved Ll, Lα, Lβ, and Lγ lines are related to the $\sigma_{Li}$ (i = 1-3) that are the ionization cross sections for the $L_1$, $L_2$, and $L_3$ as

$$\sigma_{Ll}^{x} = [\sigma_{L1}(f_{12}f_{23} + f_{13}) + \sigma_{L2}f_{23} + \sigma_{L3}]\omega_3 F_{3l} \quad (4a)$$

$$\sigma_{L\alpha}^{x} = [\sigma_{L1}(f_{12}f_{23} + f_{13}) + \sigma_{L2}f_{23} + \sigma_{L3}]\omega_3 F_{3\alpha} \quad (4b)$$

$$\sigma_{L\beta}^{x} = \sigma_{L1}[\omega_1 F_{1\beta} + f_{12}\omega_2 F_{2\beta} + (f_{12}f_{23} + f_{13})\omega_3 F_{3\beta}] + \sigma_{L2}(\omega_2 F_{2\beta} + f_{23}\omega_3 F_{3\beta}) + \sigma_{L3}\omega_3 F_{3\beta} \quad (4c)$$

$$\sigma_{L\gamma}^{x} = \sigma_{L1}(\omega_1 F_{1\gamma} + f_{12}\omega_2 F_{2\gamma}) + \sigma_{L2}\omega_2 F_{2\gamma} \quad (4d)$$

The measured L line x-ray production cross sections and the calculated ones using the $L_i$ subshell ionization cross sections from different theories including the correction for multiple ionization (MI) effects, *viz.*, FBA-MI [49–51,54], ECPSSR-MI [55], and ECUSAR-MI [56] are given in Table 1. The theoretical cross sections have been calculated using the $L_i$ subshell ionization cross sections corresponding to the incident ion charge state. A representative case of $L_i$ subshell ionization in gold bombarded by different charge states of $^{19}$F projectile ions based on the FBA [49–51,54] and ECUSAR [56] is shown in the Figure 4. $F_{ip}$ (i= 1-3, p = l, α, β, γ) are the radiative fractional emission rates. The L x-ray emission rates based on DHS calculation [57] and the interpolated values by Campbell and Wang [45] have been used in the present measurements. For the two datasets of $F_{3\beta}$, $F_{1\gamma}$, and $F_{2\gamma}$ values, the difference is 5-8% over the atomic range $Z_T$ = 50-92, whereas, other values of the emission rates differ from each other by less than 4%. The parameters $\omega_i$ (i = 1-3) are the fluorescence yields of the $L_i$ subshells and $f_{ij}$ (i<j) are the CK yields for the transition between $L_i$ and $L_j$ subshells. The single-hole fluorescence $\omega_i^0$ and CK yields $f_{ij}^0$ can be obtained from Krause [58] and Chen *et al.* [59]. As given in Table 2, the datasets of $\omega_i^0$ and $f_{ij}^0$ significantly differ from each other. For the present elements under consideration, the $f_{13}^0$ (Rec.) values are on the average about 15% lower than the $f_{13}^0$ (DHS) values and ~ 9% higher than the $f_{13}^0$ (Krause) values. The $f_{12}^0$(Rec.) values differ ~ 15% in average higher than the $f_{12}^0$ (DHS) values and are about half the $f_{12}^0$(Krause) values. The $f_{23}^0$(Rec.) values from different sets do not differ significantly. The $\omega_2^0$(Rec.) and $\omega_3^0$(Rec.) values agree with the DHS values and are higher from Krause's values below 10% for the present elements. The $\omega_1^0$(Rec.) values differ from $\omega_1^0$(Krause) values by 0-14% and from $\omega_1^0$(DHS) by 13-52%. The use of different sets of atomic parameters can change x-ray production cross section by ~30%. Recent values of $\omega_i^0$ and $f_{ij}^0$ compiled by Campbell [43,44] for the elements with $25 \leq Z \leq 96$ have been used in the present work for singly-ionized atoms.

## 4. EFFECT OF SINGLE- AND MULTIPLE-HOLE ATOMIC PARAMETERS ON THE CONVERSION OF IONIZATION TO X-RAY PRODUCTION CROSS SECTIONS

Multiple vacancies in the target atom change the atomic parameters by increasing fluorescence yields and decreasing CK yields which in turn enhances x-ray production cross sections. In the present work, single-hole fluorescence $\omega_i^0$ and CK yields $f_{ij}^0$ [43], were corrected for multiple ionization using a model prescribed by Lapicki *et al.* [60]. Each electron in a manifold of the outer subshells is ionized with a probability P which is calculated from Equation (A3) of [60] and replacing the projectile atomic number $Z_P$ by its charge state q [61],

$$P = \frac{q^2}{2\beta v_P^2}\left(1 - \frac{\beta}{4v_P^2}\right) \qquad (5)$$

With $\beta$ = 0.9. For charge state q, we take the incident charge state of the projectile. The $\omega_i^0$ values corrected for simultaneous ionization in outer subshells are given by

$$\omega_i = \omega_i^o\left[1 - P(1 - \omega_i^o)\right]^{-1} \qquad (6)$$

While the $f_{ij}$ values for multiple ionization are given by

$$f_{ij} = f_{ij}^o(1 - P)^2. \qquad (7)$$

Note that the fractional rates $F_{ip}$ remain unchanged because both partial and total non-radiative widths are narrowed by identical factors. With Eq. (6) and Eq. (7), the single-hole fluorescence and CK yields change at different ion beam energies and charge states. Fluorescence and CK yields for singly- and multiply-ionized $_{78}$Pt and $_{92}$U elements are given in Table 3. It is clear from this table that in the extreme the $L_i$ subshell fluorescence yields are enhanced by ~ 15% and CK yields are reduced up to ~27% from single-hole to multiple-hole atom in $_{78}$Pt. These values differ by 2-3% over the range of the ion beam energies and the projectile charge states used in the present experiment.

L-shell line and total x-ray production cross sections, at corresponding energies and incident charge state of the fluorine ions, are listed in Table 1 and shown in Figures 5-7. Although the connection between observed L x-ray lines and calculated $L_i$ subshell ionization cross sections depends on a combination of intra-shell coupling and inner shell multiple ionization effects [62,63], the data for ionization of comparably heavy targets as ours but by significantly slower 4-8 MeV carbon ions [64] show that multiple-ionization is more effective. While both effects subside in the 4-6 MeV/amu range of the present experiment, the effect of the intra-shell coupling is overshadowed by multiple ionization [64]. Thus in Table 1, ignoring the negligible effect the intra-shell coupling, all measured cross sections are compared to the predictions of the FBA [49–51,54], ECPSSR [55], and ECUSAR [56] ionization theories converted to the x-ray production cross sections using multiple-hole atomic parameters calculated with Eqs.(5)-(7). Within experimental error, the ratios of our data to ECPSSR [55] and ECUSAR [56] are practically the same at q = 6$^+$ and q = 7$^+$; at q = 8$^+$, ECUSAR [56] is distinctly better than ECPSSR[55]. After averaging over energies and charge states, for each element Table 4 shows the standard deviations of the so calculated x-ray production cross sections from our measurements. For comparison this table also shows the standard deviations when the ionization theories are converted with single-hole atomic parameters that are listed in Table 3. Theories converted with multiply-ionized atomic parameters are clearly in better agreement with our data.

## 5. CONCLUSIONS

In the present work, the L x-ray production cross sections of $_{78}$Pt, $_{79}$Au, $_{82}$Pb, $_{83}$Bi, $_{90}$Th, and $_{92}$U elements for the incident $^{19}$F ions of charge states 6$^+$, 7$^+$ and 8$^+$ have been measured. These data were compared with the theoretical L x-ray production cross section calculated from the $L_i$ ($i$ = 1-3) subshell ionization cross sections using FBA, ECPSSR and ECUSAR and recently recommended set of the $L_i$ ($i$ = 1-3) subshell fluorescence and CK yields with and without modifications for the multiple vacancies in the outer shells. While the measured values are about two times lower than those calculated using the FBA, exhibit agreement with those based on the ECPSSR and ECUSAR calculations. This is particularly so when the fluorescence yields are corrected for the outer-shell multiple ionization. Although the ionization cross sections for the $^{19}$F ions with the $6^+$, $7^+$, and $8^+$ charge states over the ion beam energies used in the present work are almost independent of the charge state, the multiple ionization effect is essentially equal in the ECPSSR-MI and ECUSAR-MI calculations for q = $6^+$ and $7^+$. At q = $8^+$, the ECUSAR-MI agrees better with the data than the ECPSSR-MI theory.

Singh et al.[28] reported L-x ray production cross sections in gold and bismuth with fluorine ions at 83 and 98 MeV of essentially the same charge state as the present data. While for $_{79}$Au their total cross sections fluctuate from as much as 7% above ours at 84 MeV to as much as 27% below ours at 98 MeV, for $_{83}$Bi their measurements are below ours by about 22% at 84 MeV and almost 43% below at 98 MeV. Even with a conservative estimate of 20% for experimental errors, the discrepancies between the Singh et al.[28] and our measurements are difficult to explain, and suggest that - aside from its inherent interest for comparison with theories and PIXE applications - it would be worth for other experimentalists to revisit this collision regime.

**Acknowledgements**

Financial support from the Science and Engineering Research Board (SERB), New Delhi to Dr. Sunil Kumar in terms of Young Scientist scheme and grant from IUAC, New Delhi in terms of UF-UP-43302 project is highly acknowledged. Author also acknowledges the work of Pelletron staff for smooth conduct of experiment.

**Figure Captions**

**Fig.1** L x-ray spectra from $_{78}$Pt, $_{79}$Au, $_{82}$Pb, $_{83}$Bi, ThF$_4$, UF$_4$ bombarded with the 98 MeV $^{19}$F ions.

**Fig.2** L x-ray spectra from $_{92}$U (48.6 μg/cm$^2$ UF$_4$ target) bombarded with 76, 84, 90, 98, 106, and 114 MeV $^{19}$F ions,

**Fig.3** Efficiency curve obtained by measuring the K x-rays fluorescence yields from targets excited by the 59.54 keV γ-ray photons. Measured values were normalized to absolute efficiency obtained using the calibrated $^{137}$Cs and $^{155}$Eu radioactive sources.

**Fig.4** L$_i$ subshell ionization in gold bombarded by $^{19}$F ions based on the FBA [49–51,54] and ECUSAR [56].

**Fig.5** Ll, Lγ, Lβ, Lα and total L x-ray production in $_{78}$Pt and $_{79}$Au targets bombarded by $^{19}$F ions according to FBA-MI, ECUSAR-MI, and present measurements.

**Fig.6** Ll, Lγ, Lβ, Lα and total L x-ray production in $_{82}$Pb and $_{83}$Bi targets bombarded by $^{19}$F ions according to FBA-MI, ECUSAR-MI, and present measurements.

**Fig.7** Ll, Lγ, Lβ, Lα and total L x-ray production in $_{90}$Th and $_{92}$U targets bombarded by $^{19}$F ions according to FBA-MI, ECUSAR-MI, and present measurements.

**Table 1.** The Ll, Lα, Lβ, Lγ, and total L x-ray production cross section (barn) in elements with 78≤$Z_T$≤92 for incident $^{19}$F ions as measured and calculated with ionization cross sections according to the FBA [49–51,54], ECPSSR [55], and ECUSAR [56] converted to x-ray production cross sections with atomic parameters modified for multiply-ionized (MI) elements [60]. The ratios of the measured to calculated cross sections are listed in the parenthesis. In bold print are the best ratios. With q = 6$^+$ and 7$^+$, the ratios of the data to ECUSAR-MI and ECPSSR-MI are (within 15% uncertainties of our measurements) statistically similar, while the ECUSAR-MI are definitely in closer agreement with the measurements than ECPSSR-MI above 100 MeV and q = 8$^+$.

| Element | $^{19}$F ion beam | | x-ray production cross sections (barn) | | | |
|---|---|---|---|---|---|---|
| | Energy (MeV) | Charge q | Measured | ECUSAR-MI | ECPSSR-MI | FBA-MI |
| $_{78}$Pt | | | | | | |
| Ll x-ray | 76 | 6$^+$ | 206 | 207(**1.00**) | 203(1.01) | 417(0.49) |
| | 84 | 6$^+$ | 226 | 253(0.89) | 249(**0.91**) | 474(0.48) |
| | 90 | 7$^+$ | 260 | 296(0.88) | 291(**0.89**) | 534(0.49) |
| | 98 | 8$^+$ | 323 | 372(0.87) | 347(**0.93**) | 678(0.48) |
| | 106 | 8$^+$ | 423 | 425(**1.00**) | 385(1.10) | 737(0.57) |
| | 114 | 8$^+$ | 416 | 477(0.87) | 448(**0.93**) | 794(0.52) |
| Lα x-ray | | | | | | |
| | 76 | 6$^+$ | 4449 | 4166(**1.07**) | 4092(1.09) | 8411(0.53) |
| | 84 | 6$^+$ | 4685 | 5110(0.92) | 5018(**0.93**) | 9567(0.49) |
| | 90 | 7$^+$ | 5749 | 5975(0.96) | 5863(**0.98**) | 10768(0.53) |
| | 98 | 8$^+$ | 7015 | 7507(0.93) | 7009(**1.00**) | 13677(0.51) |
| | 106 | 8$^+$ | 8602 | 8568(**1.00**) | 7758(1.11) | 14872(0.58) |
| | 114 | 8$^+$ | 8811 | 9624(**0.92**) | 8045(1.10) | 16008(0.55) |
| Lβ x-ray | | | | | | |
| | 76 | 6$^+$ | 2532 | 2373(**1.07**) | 2339(1.08) | 5169(0.49) |
| | 84 | 6$^+$ | 3037 | 2962(**1.03**) | 2919(1.04) | 5940(0.51) |
| | 90 | 7$^+$ | 3613 | 3517(**1.03**) | 3463(1.04) | 6826(0.53) |
| | 98 | 8$^+$ | 4338 | 4466(0.971) | 4219(**1.028**) | 8593(0.50) |
| | 106 | 8$^+$ | 5507 | 5169(**1.07**) | 4713(1.17) | 9482(0.58) |
| | 114 | 8$^+$ | 6007 | 5879(**1.02**) | 5576(1.08) | 10340(0.58) |
| Lγ x-ray | | | | | | |
| | 76 | 6$^+$ | 346 | 342(**1.01**) | 337(1.03) | 776(0.45) |
| | 84 | 6$^+$ | 498 | 433(**1.15**) | 428(1.16) | 898(0.55) |

|          |     |       |             |             |             |
|----------|-----|-------|-------------|-------------|-------------|
|          | 90  | 7⁺    | 481         | 521(0.92)   | 514(**0.94**) | 1051(0.46)  |
|          | 98  | 8⁺    | 598         | 667(0.90)   | 636(**0.94**) | 1315(0.45)  |
|          | 106 | 8⁺    | 804         | 780(**1.03**) | 715(1.12)   | 1466(0.55)  |
|          | 114 | 8⁺    | 883         | 895(**0.99**) | 855(1.03)   | 1615(0.55)  |
| Total L  |     |       |             |             |             |             |
|          | 76  | 6⁺    | 7533        | 7087(**1.06**) | 6971(1.08) | 14773(0.51) |
|          | 84  | 6⁺    | 8377        | 8758(0.96)  | 8614(**0.97**) | 16880(0.50) |
|          | 90  | 7⁺    | 10103       | 10309(0.98) | 10131(**1.00**) | 19180(0.53) |
|          | 98  | 8⁺    | 12274       | 13012(0.94) | 12211(**1.01**) | 24263(0.51) |
|          | 106 | 8⁺    | 15337       | 14941(**1.03**) | 13571(1.13) | 26558(0.58) |
|          | 114 | 8⁺    | 16117       | 16876(0.96) | 15924(**1.01**) | 28755(0.56) |

$_{79}$Au

Ll x-ray

|         |     |    |      |            |            |            |
|---------|-----|----|------|------------|------------|------------|
|         | 76  | 6⁺ | 182  | 193(0.94)  | 189(**0.96**) | 395(0.46)  |
|         | 84  | 6⁺ | 220  | 237(0.93)  | 233(**0.94**) | 450(0.49)  |
|         | 90  | 7⁺ | 270  | 278(0.97)  | 273(**0.99**) | 508(0.53)  |
|         | 98  | 8⁺ | 323  | 349(0.93)  | 327(**0.99**) | 643(0.50)  |
|         | 106 | 8⁺ | 453  | 400(**1.13**) | 363(1.25)  | 701(0.65)  |
|         | 114 | 8⁺ | 463  | 451(**1.03**) | 424(1.09)  | 757(0.61)  |

Lα x-ray

|         |     |    |      |            |            |            |
|---------|-----|----|------|------------|------------|------------|
|         | 76  | 6⁺ | 3854 | 3829(**1.01**) | 3765(1.02) | 7849(0.49) |
|         | 84  | 6⁺ | 3921 | 4713(0.83) | 4634(**0.85**) | 8952(0.44) |
|         | 90  | 7⁺ | 5344 | 5521(0.97) | 5421(**0.99**) | 10097(0.53) |
|         | 98  | 8⁺ | 6416 | 6939(0.93) | 6498(**0.99**) | 12775(0.50) |
|         | 106 | 8⁺ | 8632 | 7948(**1.09**) | 7210(1.20) | 13940(0.62) |
|         | 114 | 8⁺ | 9574 | 8958(**1.07**) | 8437(1.13) | 15054(0.64) |

Lβ x-ray

|         |     |    |      |            |            |            |
|---------|-----|----|------|------------|------------|------------|
|         | 76  | 6⁺ | 2169 | 2160(**1.00**) | 2130(1.03) | 4776(0.45) |
|         | 84  | 6⁺ | 2397 | 2704(0.89) | 2667(**0.90**) | 5504(0.44) |
|         | 90  | 7⁺ | 3265 | 3216(**1.02**) | 3168(1.03) | 6333(0.52) |

|     |     |       |             |             |             |
| --- | --- | ----- | ----------- | ----------- | ----------- |
| 98  | 8⁺  | 3850  | 4084(0.94)  | 3868(**1.00**) | 7945(0.48)  |
| 106 | 8⁺  | 5382  | 4742(**1.13**) | 4331(1.24) | 8792(0.61)  |
| 114 | 8⁺  | 6224  | 5411(**1.15**) | 5141(1.21) | 9614(0.65)  |

Lγ x-ray

|     |     |     |             |             |             |
| --- | --- | --- | ----------- | ----------- | ----------- |
| 76  | 6⁺  | 294 | 313(0.94)   | 309(**0.95**) | 723(0.41)   |
| 84  | 6⁺  | 276 | 398(0.69)   | 393(**0.70**) | 839(0.33)   |
| 90  | 7⁺  | 441 | 479(0.92)   | 473(**0.93**) | 983(0.45)   |
| 98  | 8⁺  | 720 | 614(**1.17**) | 587(1.23)   | 1225(0.59)  |
| 106 | 8⁺  | 791 | 721(**1.10**) | 662(1.19)   | 1371(0.58)  |
| 114 | 8⁺  | 987 | 830(**1.19**) | 794(1.24)   | 1513(0.65)  |

Total L

|     |     |       |             |             |             |
| --- | --- | ----- | ----------- | ----------- | ----------- |
| 76  | 6⁺  | 6499  | 6494(**1.00**) | 6394(1.02) | 13742(0.47) |
| 84  | 6⁺  | 6814  | 8052(0.85)  | 7928(**0.86**) | 15746(0.43) |
| 90  | 7⁺  | 9319  | 9493(0.98)  | 9335(**1.00**) | 17920(0.52) |
| 98  | 8⁺  | 11309 | 11987(0.94) | 11279(**1.00**) | 22587(0.50) |
| 106 | 8⁺  | 15258 | 13811(1.14) | 12565(1.21) | 24804(0.62) |
| 114 | 8⁺  | 17249 | 15649(1.10) | 14797(1.17) | 26938(0.64) |

$_{82}$Pb

Ll x-ray

|     |     |     |             |             |             |
| --- | --- | --- | ----------- | ----------- | ----------- |
| 76  | 6⁺  | 152 | 138(1.10)   | 151(**1.01**) | 327(0.46)   |
| 84  | 6⁺  | 163 | 189(0.86)   | 187(**0.87**) | 375(0.43)   |
| 90  | 7⁺  | 215 | 223(0.96)   | 219(**0.98**) | 425(0.51)   |
| 98  | 8⁺  | 260 | 280(0.93)   | 264(**0.98**) | 533(0.49)   |
| 106 | 8⁺  | 370 | 324(**1.14**) | 295(1.25)   | 586(0.63)   |
| 114 | 8⁺  | 325 | 367(0.89)   | 348(**0.93**) | 637(0.51)   |

Lα x-ray

|     |     |      |             |             |             |
| --- | --- | ---- | ----------- | ----------- | ----------- |
| 76  | 6⁺  | 3010 | 2611(1.15)  | 2858(**1.05**) | 6206(0.49)  |
| 84  | 6⁺  | 3000 | 3595(0.83)  | 3539(**0.85**) | 7122(0.42)  |
| 90  | 7⁺  | 4202 | 4228(**0.994**) | 4159(1.010) | 8057(0.52)  |
| 98  | 8⁺  | 4794 | 5321(0.90)  | 5015(**0.96**) | 10110(0.47) |
| 106 | 8⁺  | 6747 | 6140(**1.10**) | 5593(1.21) | 11110(0.61) |
| 114 | 8⁺  | 6799 | 6971(**0.98**) | 6596(1.03) | 12077(0.56) |

Lβ x-ray

|  |  |  |  |  |  |  |
|---|---|---|---|---|---|---|
|  | 76 | 6+ | 1662 | 1559(**1.07**) | 1605(**1.04**) | 3752(0.44) |
|  | 84 | 6+ | 1743 | 2048(0.85) | 2022(**0.86**) | 4354(0.40) |
|  | 90 | 7+ | 2455 | 2445(**1.00**) | 2412(1.02) | 5021(0.49) |
|  | 98 | 8+ | 2886 | 3108(0.93) | 2961(**0.97**) | 6247(0.46) |
|  | 106 | 8+ | 4112 | 3637(**1.13**) | 3334(1.23) | 6962(0.59) |
|  | 114 | 8+ | 4415 | 4181(**1.06**) | 3993(1.11) | 7665(0.58) |
| Lγ x-ray |  |  |  |  |  |  |
|  | 76 | 6+ | 216 | 236(0.915) | 234(**0.923**) | 572(0.38) |
|  | 84 | 6+ | 212 | 302(0.70) | 299(**0.71**) | 670(0.32) |
|  | 90 | 7+ | 371 | 366(**1.01**) | 362(1.02) | 787(0.47) |
|  | 98 | 8+ | 398 | 470(0.85) | 451(**0.88**) | 973(0.41) |
|  | 106 | 8+ | 587 | 556(**1.06**) | 512(1.15) | 1097(0.54) |
|  | 114 | 8+ | 647 | 646(**1.00**) | 622(1.04) | 1220(0.53) |
| Total L |  |  |  |  |  |  |
|  | 76 | 6+ | 5039 | 4544(1.11) | 4848(**1.04**) | 10858(0.46) |
|  | 84 | 6+ | 5118 | 6135(0.83) | 6047(**0.85**) | 12522(0.41) |
|  | 90 | 7+ | 7242 | 7262(**1.00**) | 7152(1.01) | 14290(0.51) |
|  | 98 | 8+ | 8338 | 9180(0.91) | 8692(**0.96**) | 17862(0.47) |
|  | 106 | 8+ | 11816 | 10658(**1.11**) | 9734(1.21) | 19754(0.60) |
|  | 114 | 8+ | 12186 | 12165(**1.00**) | 11558(1.05) | 21598(0.56) |

$_{83}$Bi

Ll x-ray

|  |  |  |  |  |  |  |
|---|---|---|---|---|---|---|
|  | 76 | 6+ | 172 | 142(**1.21**) | 140(1.23) | 307(0.56) |
|  | 84 | 6+ | 162 | 176(0.92) | 173(**0.94**) | 353(0.46) |
|  | 90 | 7+ | 181 | 207(0.87) | 204(**0.89**) | 400(0.45) |
|  | 98 | 8+ | 259 | 261(**0.99**) | 246(1.05) | 500(0.52) |
|  | 106 | 8+ | 361 | 301(**1.12**) | 275(1.31) | 551(0.66) |
|  | 114 | 8+ | 311 | 343(0.91) | 325(**0.96**) | 600(0.52) |
| Lα x-ray |  |  |  |  |  |  |
|  | 76 | 6+ | 2963 | 2648(**1.12**) | 2610(1.14) | 5744(0.52) |
|  | 84 | 6+ | 2860 | 3287(0.87) | 3238(**0.88**) | 6603(0.43) |
|  | 90 | 7+ | 3428 | 3871(0.89) | 3811(**0.90**) | 7476(0.46) |
|  | 98 | 8+ | 4730 | 4873(0.971) | 4602(**1.028**) | 9355(0.51) |
|  | 106 | 8+ | 6015 | 5635(**1.07**) | 5138(1.17) | 10301(0.58) |
|  | 114 | 8+ | 6557 | 6411(**1.02**) | 6075(1.08) | 11220(0.58) |

Lβ x-ray

| | | | | | | |
|---|---|---|---|---|---|---|
| | 76 | $6^+$ | 1630 | 1480(**1.10**) | 1462(1.11) | 3465(0.47) |
| | 84 | $6^+$ | 1648 | 1868(0.88) | 1845(**0.89**) | 4028(0.41) |
| | 90 | $7^+$ | 2049 | 2232(0.92) | 2203(**0.93**) | 4648(0.44) |
| | 98 | $8^+$ | 2746 | 2839(0.97) | 2709(**1.01**) | 5767(0.48) |
| | 106 | $8^+$ | 3638 | 3329(**1.09**) | 3055(1.19) | 6441(0.56) |
| | 114 | $8^+$ | 4253 | 3836(**1.11**) | 3668(1.16) | 7106(0.60) |

Lγ x-ray

| | | | | | | |
|---|---|---|---|---|---|---|
| | 76 | $6^+$ | 213 | 215(0.99) | 213(**1.00**) | 530(0.40) |
| | 84 | $6^+$ | 190 | 276(0.688) | 274(**0.693**) | 622(0.31) |
| | 90 | $7^+$ | 255 | 335(0.76) | 331(**0.77**) | 731(0.35) |
| | 98 | $8^+$ | 374 | 430(0.87) | 414(**0.90**) | 902(0.41) |
| | 106 | $8^+$ | 657 | 511(**1.29**) | 471(1.39) | 1019(0.64) |
| | 114 | $8^+$ | 596 | 595(**1.00**) | 573(1.04) | 1137(0.52) |

Total L

| | | | | | | |
|---|---|---|---|---|---|---|
| | 76 | 6+ | 4978 | 4485(**1.11**) | 4425(1.12) | 10046(0.50) |
| | 84 | 6+ | 4859 | 5607(0.87) | 5530(**0.88**) | 11607(0.42) |
| | 90 | 7+ | 5913 | 6645(0.89) | 6550(**0.90**) | 13255(0.45) |
| | 98 | 8+ | 8109 | 8402(0.97) | 7972(**1.02**) | 16525(0.49) |
| | 106 | 8+ | 10671 | 9776(**1.09**) | 8940(1.19) | 18313(0.58) |
| | 114 | 8+ | 11717 | 11185(**1.05**) | 10642(1.10) | 20064(0.58) |

$_{90}$Th

Ll x-ray

| | | | | | | |
|---|---|---|---|---|---|---|
| | 76 | $6^+$ | 86 | 85(**1.01**) | 84(1.02) | 203(0.42) |
| | 84 | $6^+$ | 108 | 106(**1.02**) | 105(1.03) | 235(0.46) |
| | 90 | $7^+$ | 148 | 126(**1.17**) | 125(1.18) | 267(0.55) |
| | 98 | $8^+$ | 144 | 159(0.91) | 152(**0.95**) | 328(0.44) |
| | 106 | $8^+$ | 204 | 186(**1.10**) | 171(1.19) | 366(0.56) |
| | 114 | $8^+$ | 297 | 213(**1.39**) | 205(1.45) | 403(0.74) |

Lα x-ray

| | | | | | | |
|---|---|---|---|---|---|---|
| | 76 | $6^+$ | 1365 | 1416(0.96) | 1401(**0.97**) | 3378(0.40) |
| | 84 | $6^+$ | 1510 | 1775(0.85) | 1755(**0.86**) | 3923(0.38) |
| | 90 | $7^+$ | 2411 | 2104(**1.15**) | 2078(1.16) | 4457(0.54) |
| | 98 | $8^+$ | 2453 | 2646(0.93) | 2534(**0.97**) | 5470(0.45) |
| | 106 | $8^+$ | 3375 | 3099(**1.09**) | 2854(1.18) | 6095(0.55) |
| | 114 | $8^+$ | 3717 | 3554(**1.05**) | 3421(1.09) | 6716(0.55) |

Lβ x-ray

|   |   |   |   |   |   |
|---|---|---|---|---|---|
| 76 | 6$^+$ | 757 | 762(0.99) | 755(**1.00**) | 1960(0.39) |
| 84 | 6$^+$ | 807 | 970(0.83) | 961(**0.84**) | 2297(0.35) |
| 90 | 7$^+$ | 1313 | 1165(**1.13**) | 1153(1.14) | 2662(0.49) |
| 98 | 8$^+$ | 1346 | 1479(0.91) | 1430(**0.94**) | 3249(0.41) |
| 106 | 8$^+$ | 1860 | 1757(**1.06**) | 1628(1.14) | 3672(0.51) |
| 114 | 8$^+$ | 2125 | 2034(**1.05**) | 1983(1.07) | 4098(0.52) |

L$\gamma$ x-ray

|   |   |   |   |   |   |
|---|---|---|---|---|---|
| 76 | 6$^+$ | 109 | 109(**1.00**) | 108(1.01) | 299(0.36) |
| 84 | 6$^+$ | 114 | 141(0.809) | 140(**0.814**) | 353(0.32) |
| 90 | 7$^+$ | 200 | 172(**1.16**) | 171(1.17) | 418(0.48) |
| 98 | 8$^+$ | 192 | 221(0.87) | 216(**0.89**) | 508(0.38) |
| 106 | 8$^+$ | 302 | 267(**1.13**) | 248(1.22) | 582(0.52) |
| 114 | 8$^+$ | 340 | 311(**1.09**) | 308(1.10) | 658(0.52) |

Total L

|   |   |   |   |   |   |
|---|---|---|---|---|---|
| 76 | 6+ | 2318 | 2371(0.98) | 2348(**0.99**) | 5839(0.40) |
| 84 | 6+ | 2538 | 2993(0.85) | 2961(**0.86**) | 6808(0.37) |
| 90 | 7+ | 4072 | 3566(**1.14**) | 3527(1.15) | 7804(0.52) |
| 98 | 8+ | 4136 | 4505(0.92) | 4332(**0.95**) | 9556(0.43) |
| 106 | 8+ | 5742 | 5309(**1.08**) | 4901(1.17) | 10715(0.54) |
| 114 | 8+ | 6479 | 6112(**1.06**) | 5918(1.09) | 11876(0.55) |

$_{92}$U

Ll x-ray

|   |   |   |   |   |   |
|---|---|---|---|---|---|
| 76 | 6$^+$ | 97 | 75(**1.29**) | 74(1.31) | 183(0.53) |
| 84 | 6$^+$ | 91 | 94(0.97) | 93(**0.98**) | 213(0.43) |
| 90 | 7$^+$ | 194 | 111(**1.75**) | 110(1.76) | 242(0.80) |
| 98 | 8$^+$ | 203 | 140(**1.45**) | 134(1.51) | 296(0.69) |
| 106 | 8$^+$ | 262 | 164(**1.60**) | 152(1.72) | 330(0.79) |
| 114 | 8$^+$ | 293 | 190(**1.54**) | 182(1.61) | 365(0.80) |

L$\alpha$ x-ray

|   |   |   |   |   |   |
|---|---|---|---|---|---|
| 76 | 6$^+$ | 1113 | 1205(0.92) | 1192(**0.93**) | 2955(0.38) |
| 84 | 6$^+$ | 1250 | 1513(0.83) | 1497(**0.84**) | 3439(0.36) |
| 90 | 7$^+$ | 1997 | 1795(**1.11**) | 1774(1.13) | 3906(0.51) |
| 98 | 8$^+$ | 2201 | 2255(0.976) | 2168(**1.015**) | 4767(0.46) |
| 106 | 8$^+$ | 2871 | 2649(**1.08**) | 2446(1.17) | 5325(0.54) |
| 114 | 8$^+$ | 3184 | 3060(**1.04**) | 2942(1.08) | 5883(0.54) |

L$\beta$ x-ray

|  | 76 | 6+ | 603 | 625(0.965) | 620(**0.973**) | 1650(0.37) |
|  | 84 | 6+ | 622 | 798(0.78) | 791(**0.79**) | 1944(0.32) |
|  | 90 | 7+ | 1084 | 959(**1.13**) | 950(1.14) | 2244(0.48) |
|  | 98 | 8+ | 1215 | 1217(**1.00**) | 1180(1.03) | 2730(0.45) |
|  | 106 | 8+ | 1605 | 1450(**1.11**) | 1345(1.19) | 3091(0.52) |
|  | 114 | 8+ | 1758 | 1697(**1.04**) | 1645(1.07) | 3459(0.51) |
| L$\gamma$ x-ray |  |  |  |  |  |  |
|  | 76 | 6+ | 79 | 86(**0.92**) | 86(**0.92**) | 244(0.32) |
|  | 84 | 6+ | 73 | 112(0.65) | 111(**0.66**) | 291(0.25) |
|  | 90 | 7+ | 174 | 137(**1.27**) | 136(1.28) | 342(0.51) |
|  | 98 | 8+ | 179 | 176(**1.02**) | 172(1.04) | 416(0.43) |
|  | 106 | 8+ | 246 | 213(**1.15**) | 199(1.24) | 478(0.51) |
|  | 114 | 8+ | 275 | 254(**1.08**) | 248(1.11) | 542(0.51) |
| Total L |  |  |  |  |  |  |
|  | 76 | 6+ | 1893 | 1991(0.95) | 1972(**0.96**) | 5032(0.38) |
|  | 84 | 6+ | 2036 | 2517(0.81) | 2492(**0.82**) | 5887(0.35) |
|  | 90 | 7+ | 3450 | 3003(**1.15**) | 2970(1.16) | 6734(0.51) |
|  | 98 | 8+ | 3798 | 3789(**1.00**) | 3654(1.04) | 8208(0.46) |
|  | 106 | 8+ | 4984 | 4476(**1.11**) | 4141(1.20) | 9224(0.54) |
|  | 114 | 8+ | 5509 | 5201(**1.06**) | 5017(1.10) | 10249(0.54) |

**Table 2.** Fluorescence and CK yields for singly-ionized elements. Values listed as recommended by Campbell [43,44] are used in the present work.

| Element | Fluorescence yield | | | | | | | | |
|---|---|---|---|---|---|---|---|---|---|
|  | $\omega_1^0$ | | | $\omega_2^0$ | | | $\omega_3^0$ | | |
|  | Campbell [43,44] | Krause [58] | Chen et al [59] | Campbell [43,44] | Krause [58] | Chen et al [59] | Campbell [43,44] | Krause [58] | Chen et al [59] |
| $_{78}$Pt | 0.114 | 0.114 | 0.074 | 0.344 | 0.321 | 0.344 | 0.303 | 0.306 | 0.303 |
| $_{79}$Au | 0.117 | 0.107 | 0.078 | 0.358 | 0.334 | 0.358 | 0.313 | 0.320 | 0.313 |
| $_{82}$Pb | 0.128 | 0.112 | 0.093 | 0.397 | 0.373 | 0.397 | 0.343 | 0.360 | 0.343 |
| $_{83}$Bi | 0.132 | 0.117 | 0.098 | 0.411 | 0.387 | 0.411 | 0.353 | 0.373 | 0.353 |
| $_{90}$Th | 0.159 | 0.161 | 0.139 | 0.503 | 0.479 | 0.503 | 0.424 | 0.463 | 0.424 |

| | | | | | | | | |
|---|---|---|---|---|---|---|---|---|
| $_{92}$U | 0.168 | 0.176 | 0.149 | 0.506 | 0.467 | 0.506 | 0.444 | 0.489 | 0.444 |

| Element | CK yield | | | | | | | | |
|---|---|---|---|---|---|---|---|---|---|
| | $f_{13}^0$ | | | $f_{12}^0$ | | | $f_{23}^0$ | | |
| | Campbell [43,44] | Krause [58] | Chen et al [59] | Campbell [43,44] | Krause [58] | Chen et al [59] | Campbell [43,44] | Krause [58] | Chen et al [59] |
| $_{78}$Pt | 0.545 | 0.500 | 0.716 | 0.075 | 0.140 | 0.067 | 0.126 | 0.124 | 0.132 |
| $_{79}$Au | 0.615 | 0.530 | 0.711 | 0.074 | 0.140 | 0.068 | 0.125 | 0.122 | 0.129 |
| $_{82}$Pb | 0.620 | 0.580 | 0.708 | 0.066 | 0.120 | 0.054 | 0.119 | 0.116 | 0.123 |
| $_{83}$Bi | 0.620 | 0.580 | 0.703 | 0.063 | 0.110 | 0.055 | 0.117 | 0.113 | 0.121 |
| $_{90}$Th | 0.620 | 0.570 | 0.659 | 0.040 | 0.090 | 0.058 | 0.103 | 0.108 | 0.106 |
| $_{92}$U | 0.620 | 0.570 | 0.660 | 0.035 | 0.080 | 0.051 | 0.140 | 0.167 | 0.139 |

**Table 3.** Fluorescence and CK yields for singly-ionized [43,44] and the ratios of atomic parameters for multiply-ionized [60] to these singly-ionized for $_{78}$Pt and $_{92}$U.

| Ion beam | | Fluorescence yield | | | CK yield | | |
|---|---|---|---|---|---|---|---|
| Energy (MeV) | Charge state q | $\omega_1$ | $\omega_2$ | $\omega_3$ | $f_{13}$ | $f_{12}$ | $f_{23}$ |
| *Singly-ionized $_{78}$Pt* | | | | | | | |
| | | 0.114 | 0.344 | 0.303 | 0.545 | 0.075 | 0.126 |
| Ratios of atomic parameters for multiply [60] to singly [43,44] ionized $_{78}$Pt | | | | | | | |
| 76 | 6$^+$ | 1.123 | 1.087 | 1.096 | 0.767 | 0.773 | 0.770 |
| 84 | 6$^+$ | 1.114 | 1.078 | 1.086 | 0.789 | 0.787 | 0.786 |
| 90 | 7$^+$ | 1.140 | 1.102 | 1.109 | 0.736 | 0.733 | 0.738 |
| 98 | 8$^+$ | 1.175 | 1.125 | 1.135 | 0.688 | 0.693 | 0.690 |
| 106 | 8⁺ | 1.167 | 1.116 | 1.122 | 0.710 | 0.707 | 0.706 |
| 114 | 8⁺ | 1.149 | 1.108 | 1.116 | 0.728 | 0.733 | 0.730 |
| Singly- ionized $_{92}$U | | | | | | | |
| | | 0.168 | 0.506 | 0.444 | 0.620 | 0.035 | 0.140 |
| Ratios of atomic parameters for multiply [60] to singly [43,44] ionized $_{92}$U | | | | | | | |
| 76 | 6⁺ | 1.11 | 1.07 | 1.07 | 0.77 | 0.77 | 0.77 |
| 84 | 6⁺ | 1.10 | 1.06 | 1.07 | 0.80 | 0.79 | 0.79 |
| 90 | 7⁺ | 1.14 | 1.08 | 1.09 | 0.74 | 0.74 | 0.74 |
| 98 | 8⁺ | 1.17 | 1.09 | 1.11 | 0.69 | 0.69 | 0.69 |
| 106 | 8⁺ | 1.15 | 1.08 | 1.10 | 0.71 | 0.71 | 0.71 |
| 114 | 8⁺ | 1.14 | 1.08 | 1.09 | 0.74 | 0.73 | 0.71 |

**Table 4.** Standard deviation of the theoretical estimates for singly (SI) and multiply (MI) ionized atoms with respect to measured values. The average standard deviation over all six target elements is also given.

| Theory | Target | Standard deviation of theories from the present experimental results | | | | |
|---|---|---|---|---|---|---|
| | | Ll | Lα | Lβ | Lγ | L-Total |
| ECPSSR-SI | Pt | 20.4 | 15.9 | 20.2 | 17.1 | 17.1 |
| | Au | 19.5 | 18.8 | 21.5 | 23.2 | 19.9 |
| | Pb | 23.0 | 18.8 | 19.8 | 15.8 | 18.9 |
| | Bi | 34.3 | 18.8 | 19.6 | 27.2 | 19.5 |
| | Th | 22.7 | 17.3 | 15.5 | 18.5 | 16.8 |
| | U | 43.7 | 16.1 | 17.6 | 20.9 | 18.2 |
| | Average | 27.3 | 17.6 | 19.0 | 20.5 | 18.4 |
| ECUSAR-SI | Pt | 18.5 | 14.6 | 22.1 | 19.9 | 17.5 |
| | Au | 16.3 | 15.9 | 19.2 | 21.4 | 17.3 |
| | Pb | 19.8 | 16.3 | 17.2 | 13.4 | 16.2 |
| | Bi | 29.8 | 15.4 | 17.0 | 24.3 | 16.3 |
| | Th | 21.6 | 14.9 | 13.3 | 16.6 | 14.6 |
| | U | 42.6 | 14.1 | 15.2 | 18.8 | 16.0 |
| | Average | 24.8 | 15.2 | 17.3 | 19.0 | 16.3 |
| ECPSSR -MI | Pt | 21.8 | 9.5 | 11.1 | 9.7 | 8.9 |
| | Au | 14.4 | 14.0 | 15.5 | 17.3 | 14.5 |
| | Pb | 19.3 | 14.2 | 13.8 | 12.3 | 13.7 |
|  | Bi | 28.5 | 13.6 | 13.7 | 23.7 | 14.0 |
|  | Th | 19.5 | 12.7 | 10.8 | 13.8 | 12.2 |
|  | U | 40.0 | 11.8 | 12.7 | 16.2 | 13.4 |
|  | Average | 23.9 | 12.6 | 12.9 | 15.5 | 12.8 |
| ECUSAR-MI | Pt | 25.3 | 10.3 | 5.1 | 8.0 | 6.1 |
|  | Au | 9.3 | 9.5 | 11.1 | 14.1 | 9.8 |
|  | Pb | 17.2 | 11.8 | 9.8 | 11.4 | 10.4 |
|  | Bi | 23.4 | 9.3 | 10.6 | 20.9 | 9.6 |
|  | Th | 17.3 | 9.6 | 8.7 | 11.9 | 9.4 |
|  | U | 37.1 | 8.5 | 9.7 | 14.0 | 10 |
|  | Average | 21.6 | 9.8 | 9.7 | 13.4 | 9.2 |

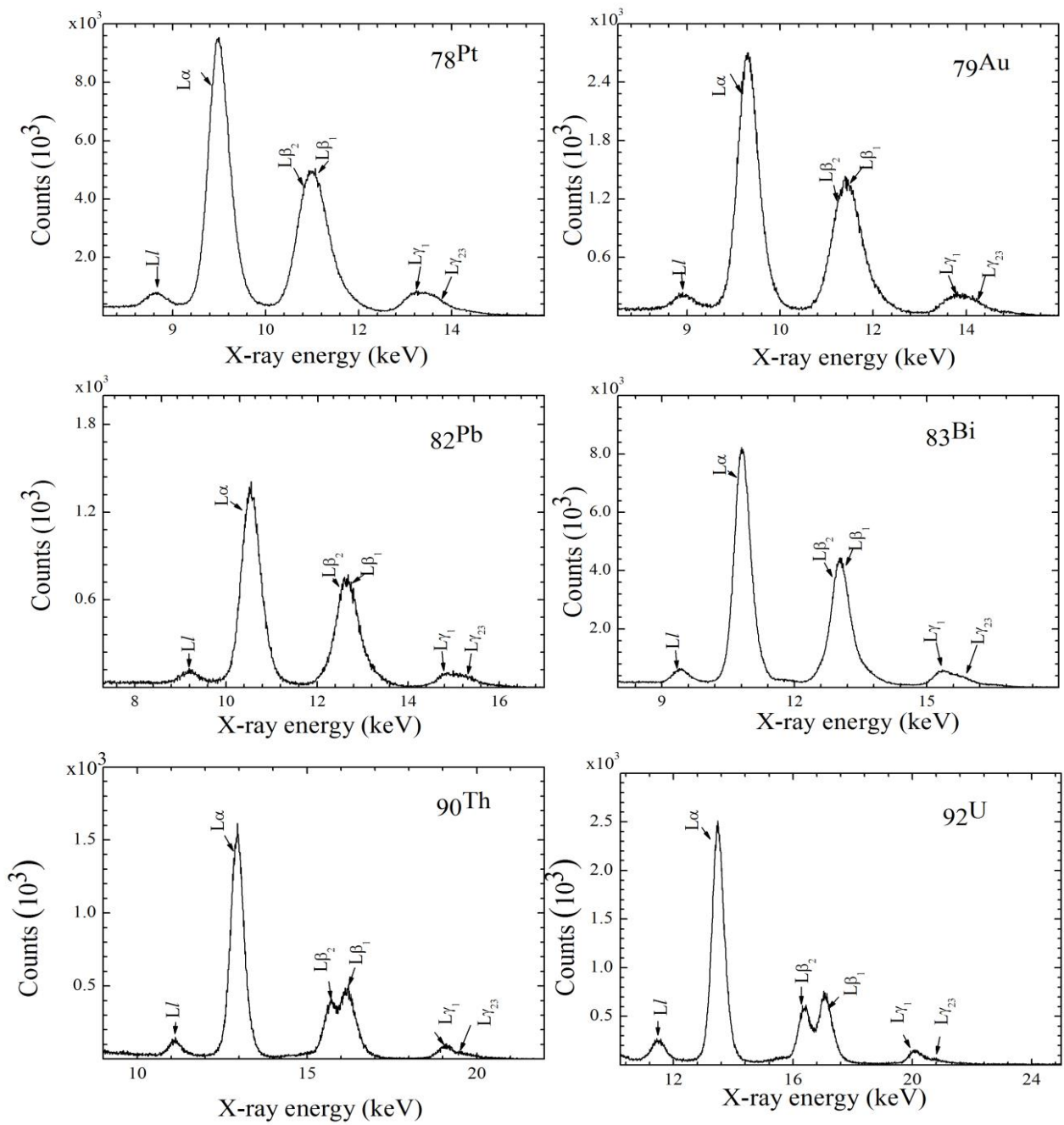

Figure 1

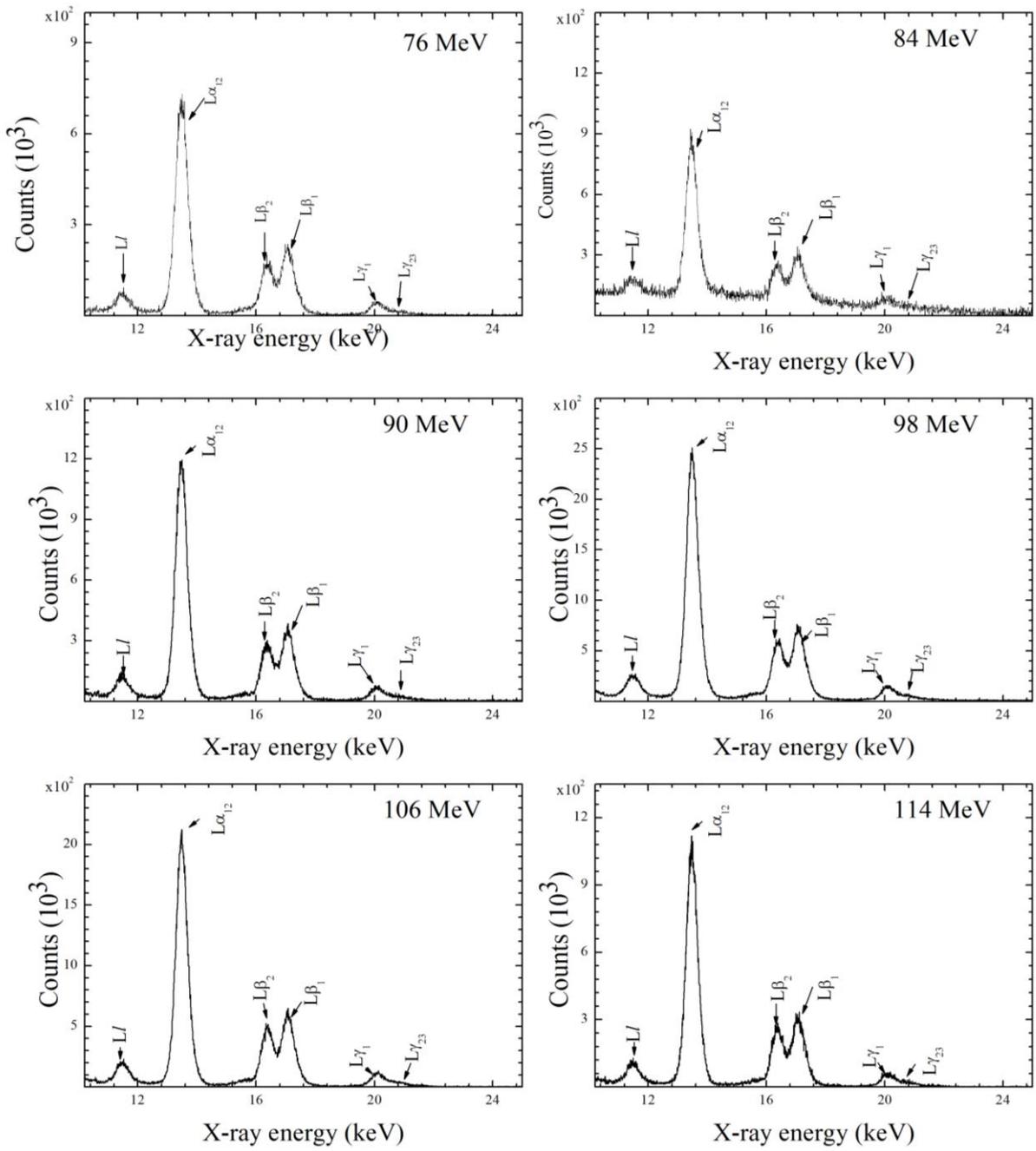

Figure 2

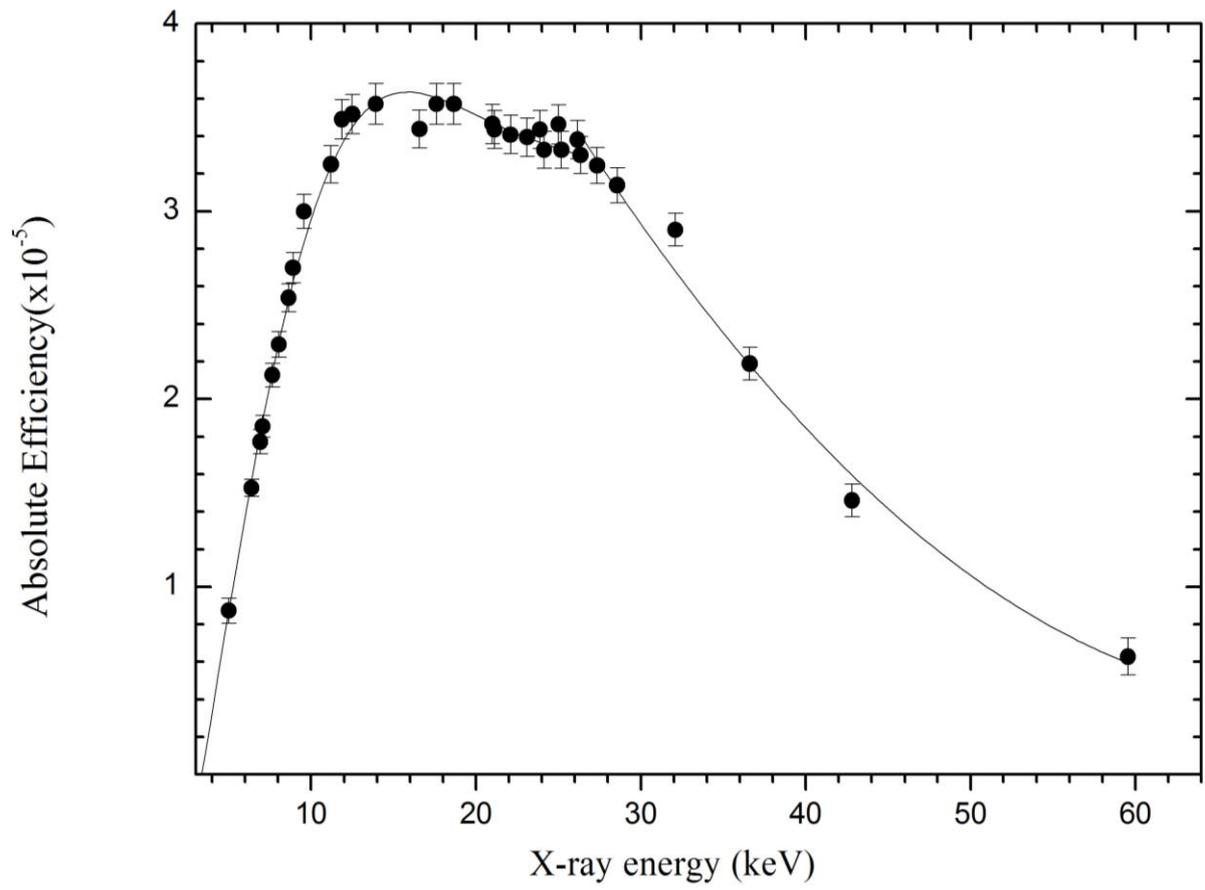

Figure 3

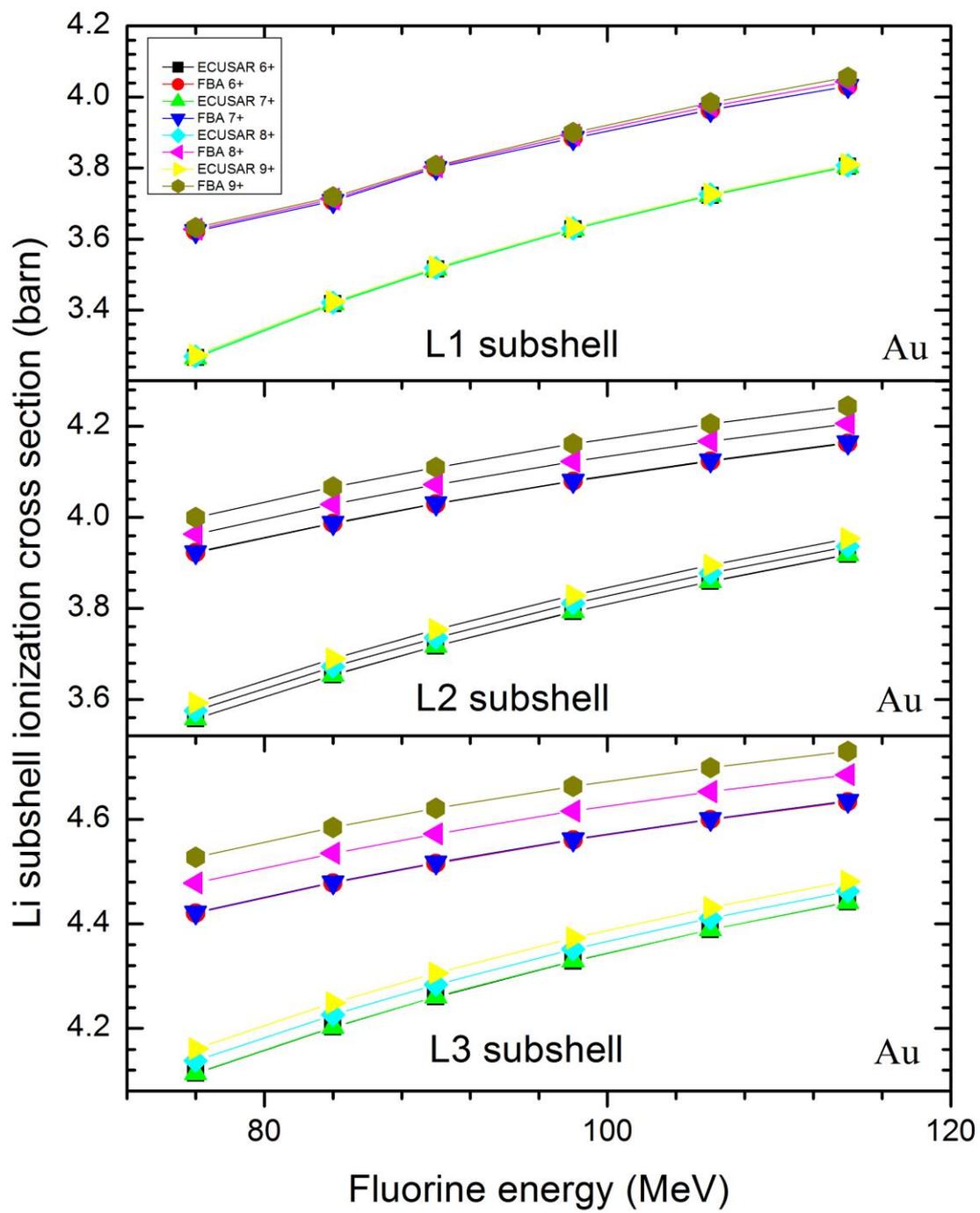

Figure 4

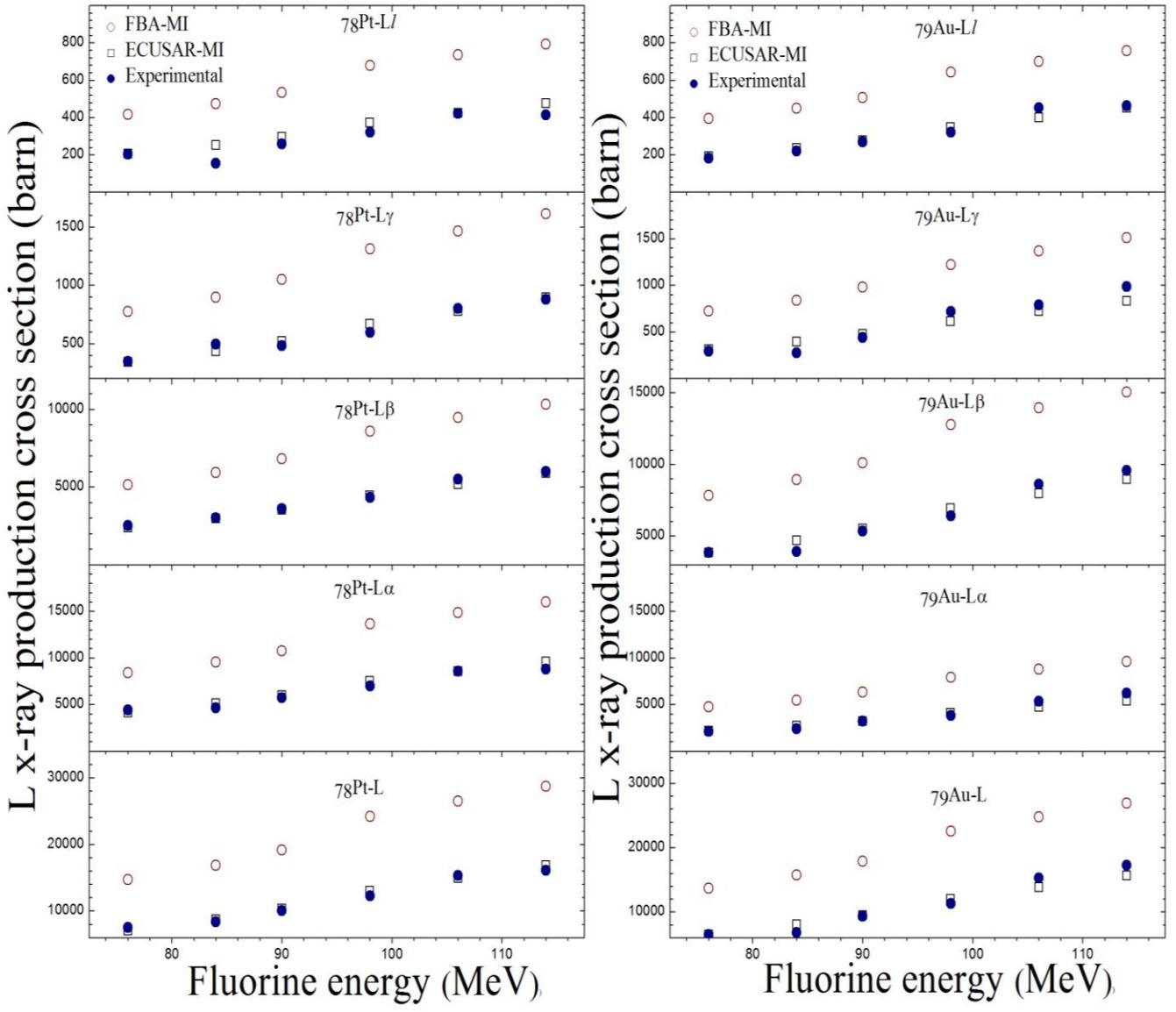

Figure 5

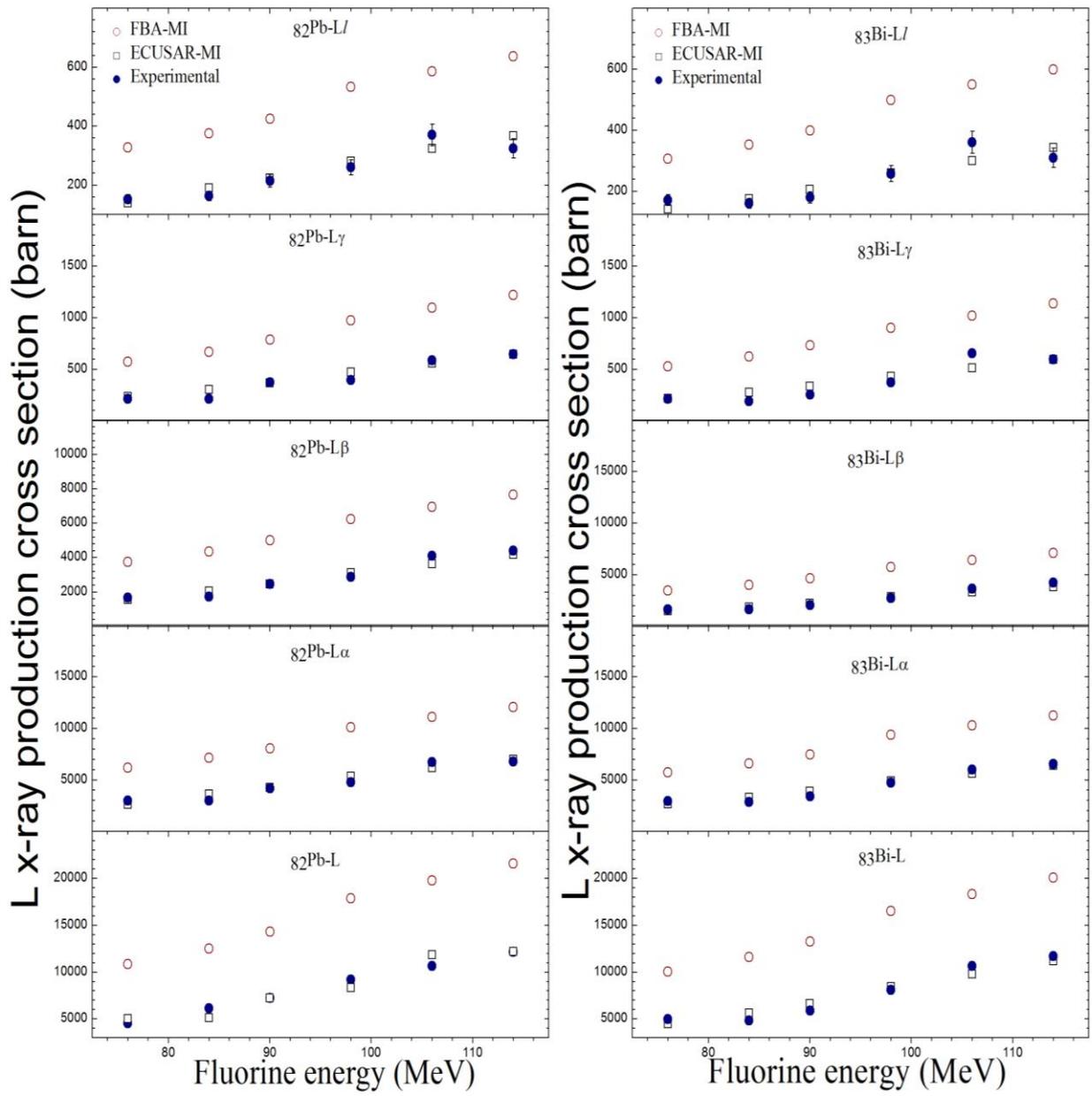

Figure 6

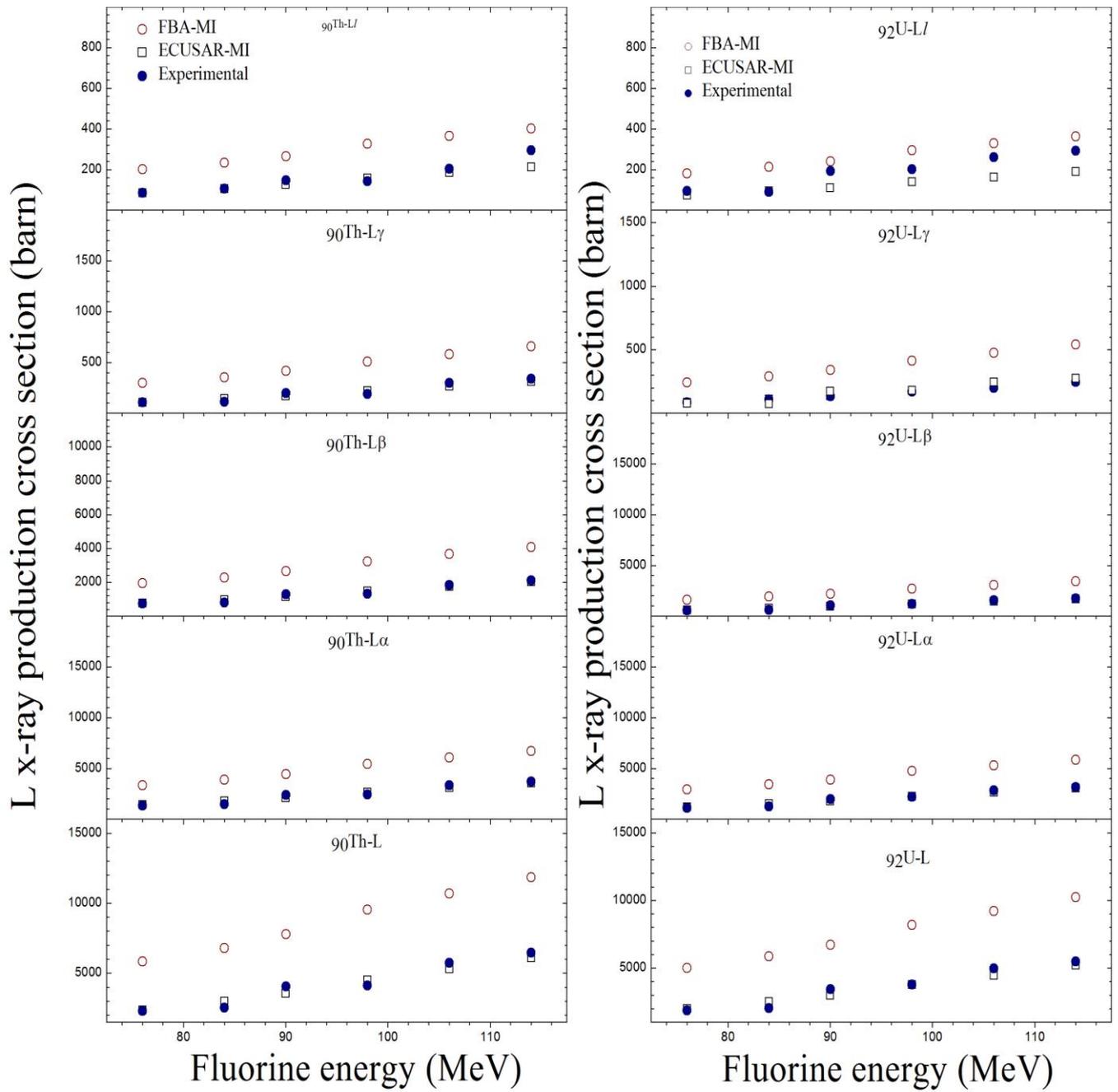

Figure 7